\documentclass[twocolumn,superscriptaddress,nofootinbib,floatfix,aps,prb,citeautoscript,longbibliography]{revtex4-2}
\usepackage[utf8]{inputenc}
\usepackage[T1]{fontenc}
\usepackage{lineno}

\usepackage{graphicx}%[draft]
\usepackage{color}
\usepackage{array}
\usepackage{dcolumn}
\usepackage{bm}
\usepackage{multirow}
\usepackage{tabularx}
\usepackage[usenames,dvipsnames]{xcolor} % Needed for comments
\usepackage{siunitx}
\usepackage{comment}
\usepackage{tikz}
\usepackage{ulem}
\usepackage{chemformula}
\let\ce\ch
\usepackage{orcidlink}
\usepackage{cleveref}
\usepackage{rotating}

\newcommand{\Tc}{\ensuremath{T_\textrm{c}}}

\newcommand{\Tcad}{\ensuremath{T_\textrm{c}^\text{AD}}}

\newcommand{\olog}{\ensuremath{\omega_\text{log}}}
\newcommand{\dosef}{\ensuremath{N_\text{F}}}

\begin{document}

\newcommand{\bochum}{Research Center Future Energy Materials and Systems of the University Alliance Ruhr and Interdisciplinary Centre for Advanced Materials Simulation, Ruhr University Bochum, Universitätsstraße 150, D-44801 Bochum, Germany}
\newcommand{\coimbra}{CFisUC, Department of Physics, University of Coimbra, Rua Larga, 3004-516 Coimbra, Portugal}
\newcommand{\mpi}{Max-Planck-Institut f\"ur Mikrostrukturphysik, Weinberg 2, D-06120 Halle, Germany}
\newcommand{\upv}{Fisika Aplikatua Saila, Gipuzkoako Ingeniaritza Eskola, University of the Basque Country (UPV/EHU), Europa Plaza 1, 20018 Donostia/San Sebastián, Spain}
\newcommand{\cfismat}{Centro de Física de Materiales (CSIC-UPV/EHU), Manuel de Lardizabal Pasealekua 5, 20018 Donostia/San Sebastián, Spain}
\newcommand{\dipc}{Donostia International Physics Center (DIPC), Manuel de Lardizabal Pasealekua 4, 20018 Donostia/San Sebastián, Spain}
\newcommand{\ludwig}{Chair for Materials Discovery and Interfaces, Institute for Materials, Ruhr University Bochum, Universitätsstraße 150, D-44801 Bochum, Germany}

%\author{Tiago F. T. Cerqueira}
\author{Tiago F. T. Cerqueira\orcidlink{0000-0002-4147-8129}}
\affiliation{\coimbra}
%\author{Yue-Wen Fang}
\author{Yue-Wen Fang\orcidlink{0000-0003-3674-7352}}
\affiliation{\upv}
\affiliation{\cfismat}
%\author{Ion Errea}
\author{Ion Errea\orcidlink{0000-0002-5719-6580}}
\affiliation{\upv}
\affiliation{\cfismat}
\affiliation{\dipc}
%\author{Antonio Sanna}
\author{Antonio Sanna\orcidlink{0000-0001-6114-9552}}
\email{sanna@mpi-halle.mpg.de}
%TEST
\affiliation{\mpi}
%\author{Miguel A. L. Marques} 
\author{Miguel A. L. Marques\orcidlink{0000-0003-0170-8222}}
\email{miguel.marques@rub.de}
\affiliation{\bochum} 

\date{\today}

\title{Searching Materials Space for Hydride Superconductors at Ambient Pressure}

\begin{abstract}
We employed a machine-learning assisted approach to search for superconducting hydrides under ambient pressure within an extensive dataset comprising over 150\,000 compounds. Our investigation yielded around 50 systems with transition temperatures surpassing 20~K, and some even reaching above 70~K. These compounds have very different crystal structures, with different dimensionality, chemical composition, stoichiometry, and arrangement of the hydrogens. Interestingly, most of these systems displayed slight thermodynamic instability, implying that their synthesis would require conditions beyond ambient equilibrium. Moreover, we found a consistent chemical composition in the majority of these systems, which combines alkali or alkali-earth elements with noble metals. This observation suggests a promising avenue for future experimental investigations into high-temperature superconductivity within hydrides at ambient pressure.

\end{abstract}

\maketitle

\section{Introduction}

A superconductor features zero electrical resistance and perfect diamagnetism when it is cooled below a critical temperature $T_c$. Such unique properties have paved the way for groundbreaking applications of superconductors, from powerful magnets in medical imaging, particle accelerators, or fusion reactors to superconducting qubits in quantum computers. Since the discovery of superconductivity in Hg in 1911, research on this field has focused on enhancing $T_c$ with particular emphasis on achieving high $T_c$ values above the boiling point of nitrogen (77~K). Nowadays, one of the primary goals in physics remains to find materials that exhibit superconductivity at ambient conditions to unlock new fundamental physics and enable wide applications of superconductors.

In the so-called unconventional superconductors, showcased by cuprate compounds, the highest superconducting transition temperatures are 134~K and 164~K at ambient pressure and 30 GPa, respectively~\cite{PhysRevB.50.4260-HKMao1994}. Nevertheless, owing to the lack of a distinct understanding of the pairing mechanism, there is at the moment no quantitative theory that can be used to predict the superconducting properties of such systems, seriously hampering the search for new unconventional compounds. On the other hand, the theoretical prediction and characterization of conventional superconductors, where the electrons at the Fermi surface are coupled by phonons to form Cooper pairs, can be well addressed by a combination of ab initio studies and Eliashberg theory. Unfortunately, since the experimental observation of superconductivity in MgB$_2$ in 2001~\cite{nagamatsu2001superconductivity}, its $T_c$ of 39~K remains the highest among the ambient-pressure conventional superconductors.

In the theory of electron-phonon coupled superconductors, the critical temperature $T_c$ is proportional to the Debye temperature $\theta_D$. Hydrogen, being the lightest element, provides a high Debye temperature to the compounds that it forms, making hydrides  the most obvious choice in the search for conventional high-$T_c$ superconductors~\cite{Ashcroft2004}. Over the past two decades, hundreds of binary hydrides, covering the majority of the chemical elements in the periodic table, have been either theoretically predicted or experimentally observed to exhibit superconductivity.
However, despite the chemically pre-compression in the hydrides, achieving high-$T_c$ superconductivity still demands substantial external pressure. For example, the experiments have suggested that $T_c > 200$~K generally requires pressures exceeding 150~GPa in the diamond anvil cell (1~GPa $\approx$ 10$^5$ atmospheric pressure): $T_c$ of 250~K in LaH$_{10}$ at 150~GPa~\cite{Hemley_LaH_PRL2019,Eremets_LaH_Nature2019}, $T_c$ of 215~K in CaH$_6$ at 172~GPa~\cite{CaH6-RRL-YanmingMA2022}, and $T_c$ of 243~K in YH$_9$ at 201~GPa~\cite{Kong-YH9-NC2021}, to name a few examples.

The search for high-$T_c$ hydrides useful for device applications has moved lately from solely prioritizing the enhancement of $T_c$ to achieving high-$T_c$ materials at low or even ambient pressure.  Unfortunately, only a few superconducting binary hydrides have been found in ambient-pressure experiments, and their critical temperatures are below or near 10~K, for instance \ce{PdH} with $T_c$ of 9~K~\cite{Stritzker1972}, \ce{TiH_{0.71}} with $T_c$ of 4.3~K~\cite{JPCM-TiH0.71-1993}, \ce{MoH_{1.2}} with $T_c$ of just 0.92~K~\cite{MoH1.2-1988Russia}, \ce{Th4H15} with $T_c$ of 8.2~K~\cite{PhysRevLett.25.741-1970-Th4H15}, \ce{NbH_{x < 0.7}} with $T_c$ of 9.4~K~\cite{NbH0.7-ZPhysikB-1977}, and \ce{ZrH3} with $T_c$ of 11.6~K~\cite{PRM2023-ZrH3-isotope}.
All these experimental studies (except the case of \ce{ZrH3}) were carried out more than 30 years ago.

Compared to the binary hydrides, the ternary, or more generally, the multinary hydrides have a much larger structural phase space, and are therefore expected to host higher $T_c$s at low and ambient pressures~\cite{Wang2021-HPSTAR-review-hyrides,BoeriRoadmap2021}. There were several experimental studies in the 1970s in ternary hydrides such as \ce{HfV2H} with a $T_c$ of 4.8~K~\cite{PRL1976-HfV2H} and \ce{Pd_{0.55}Cu_{0.45}H_{0.7}} with a $T_c$ of 16.6~K~\cite{Stritzker1974-Pd-Noble-H}. However, more recently, theoretical reports that are facilitated by high-throughput calculations, high-$T_c$ templates, crystal structure prediction, or machine learning methods, have pioneered the exploration of promising superconducting hydrides at low/ambient pressure. Due to the strong quantum ionic and anharmonic effects,  both \ce{BaSiH8}~\cite{Lucrezi2023-BaSiH8-anharmonic} and \ce{LuN4H11}~\cite{fang2023-LuNH-arxiv} are predicted to exhibit dynamical stability and high $T_c$ of 94~K and 100~K, respectively, at 20~GPa. The compound \ce{CsBH5} is predicted to have a strong electron-phonon coupling constant $\lambda$ of 3.96 at 1~GPa, resulting in a high $T_c$ of 83~K~\cite{CsBH5-1GPa-MiaoGAO-Ningbo-PRB2023}.  The deep-learning-recommended \ce{MgBH}~\cite{Choudhary2022} and the perovskite \ce{MgHCu3}~\cite{Tian2023} are predicted by electron-phonon coupling calculations to show $T_c$ values of 15.5~K and 42~K at atmospheric pressure, respectively. Finally, \ce{Mg2IrH6} with a $T_c$ above 80~K was predicted, together with other compounds with the same cubic structure, in our recent study~\cite{Sanna_2024} and in Ref.~\cite{dolui2023-Mg2IrH6-arxiv}. 

Herein, we use a machine-learning method to accelerate the search of superconductivity in a dataset of more than 150\,000 hydrides. Our workflow has identified 54 compounds with $T_c$ exceeding 20~K, most of which have not been reported previously.  Our work provides a huge platform for ambient-pressure superconducting hydrides, inviting worldwide experimental groups to explore their synthesis, and further study the transport properties, Messier effect, isotope effect and potential device applications.

\section{Results and discussion}

Here we are concerned with hydrides only. To identify such compounds that are potentially high-temperature conventional superconductors at ambient pressure we used a machine learning accelerated workflow along the lines of Ref.~\onlinecite{advmat}. Unfortunately, the dataset of Ref.~\onlinecite{advmat} did not include any high-temperature hydride (with the exception of \ce{PdH} that has a too high-\Tc\ in the harmonic approximation), so the machine trained in this dataset had difficulties predicting hydrides. To circumvent this problem, we added manually a series of hydrides to the dataset, and retrained the machine, that was subsequently used to predict the superconducting properties of the hydrides present in the \textsc{alexandria} database~\cite{sciadv,advmat}. Compounds predicted to have \Tc\ above 20~K were then analysed with density-functional perturbation theory (DFPT) and added to the dataset. The procedure was then iterated three times. Our final model had an error of 1.39~K for \Tc, 22.47~K for \olog, and 0.16 for $\lambda$. This should be compared to the average values in the training set of 3.41~K, 233.38~K, and 0.48, respectively.

In \cref{tab:propa} and \cref{tab:propb} we present all the hydrides predicted to have superconducting transition temperatures above 20~K (see Supplementary Information for the details of each compound). We note that the values presented in the tables are the ones obtained with density-functional perturbation theory and the Allen-Dynes formula, and not the values predicted by the machine-learning model. We divided the compounds by families, and ordered by \Tc, in order to facilitate the analysis. As we can see, many compounds belong to just a few families, such as the SM$_2$-TM-H$_6$ family~\cite{Sanna_2024,10.48550/arxiv.2310.07562,10.48550/arxiv.2401.17024}, perovskites (both normal and inverted), and double perovskites. There is also a striking regularity in what concerns the chemical compositions. In fact, the large majority of the hydrides include both a simple metal (SM), typically an alkali, alkali earth, or even a IIIA or IVA element, and a transition (TM) or noble metal.

\begin{figure*}
\begin{tabular}{c c c c}
  (a)~\ce{Mg2PtH6} & (b)~\ce{KInH3} & (c)~\ce{K2InCuH6}  & (d)~\ce{LiPdH2} \\
  \includegraphics[height=4cm]{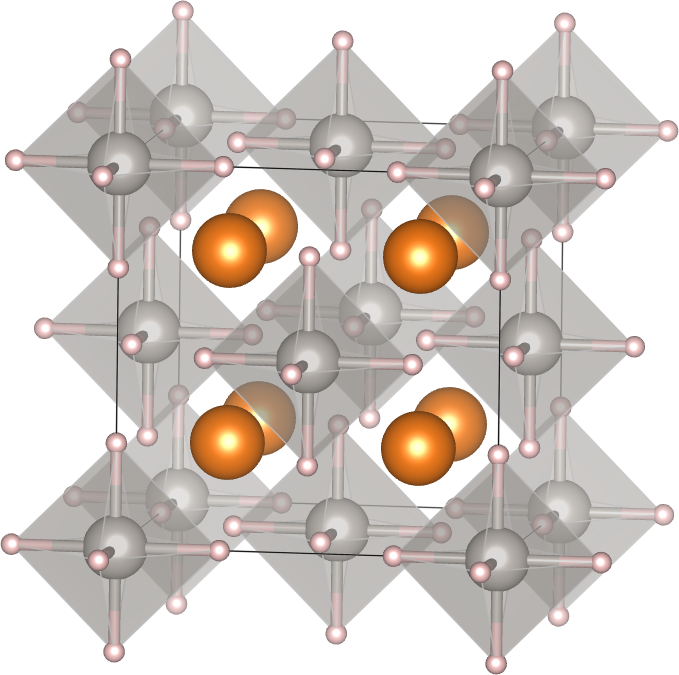} &
  \includegraphics[height=4cm]{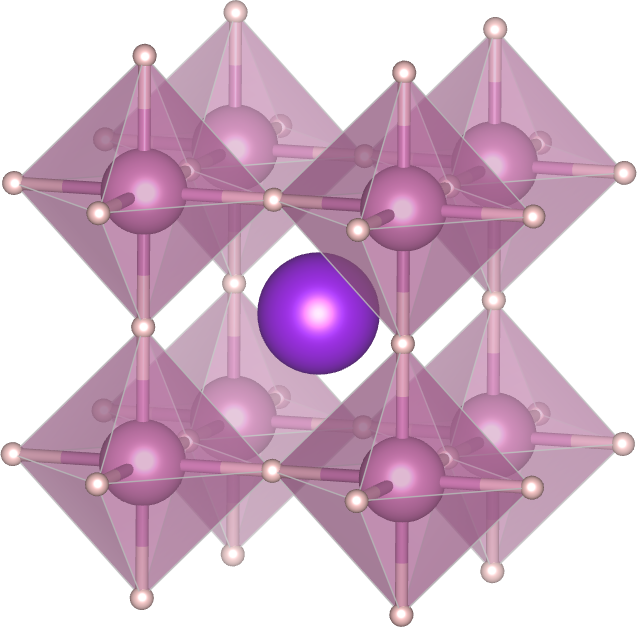} &
  \includegraphics[height=4cm]{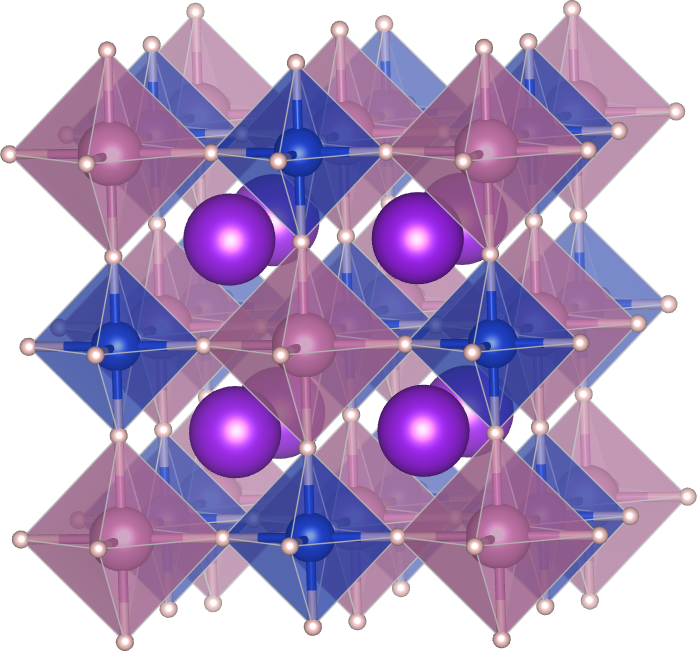} &
  \includegraphics[height=4cm]{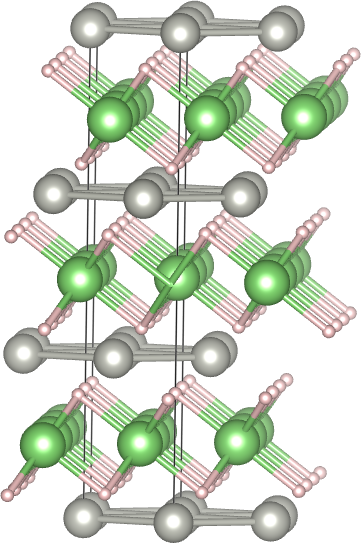}
  \\
  (e)~\ce{Na3Pd2H9} & (f)~\ce{Li2AuH2} & (g)~\ce{ZrH3} & (h)~\ce{CsBePtH6} \\
  \includegraphics[height=4cm]{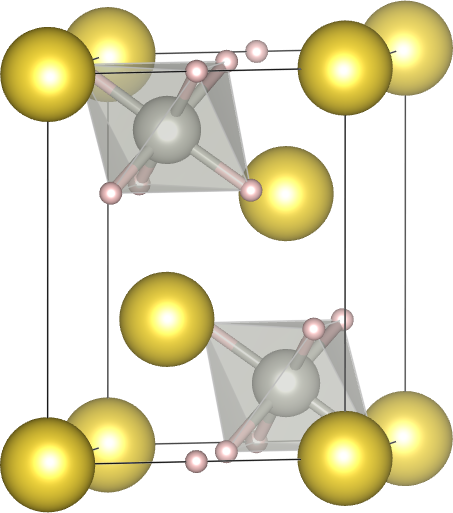} &
  \includegraphics[height=4cm]{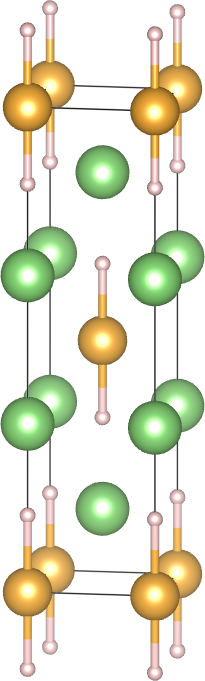} &
  \includegraphics[height=4cm]{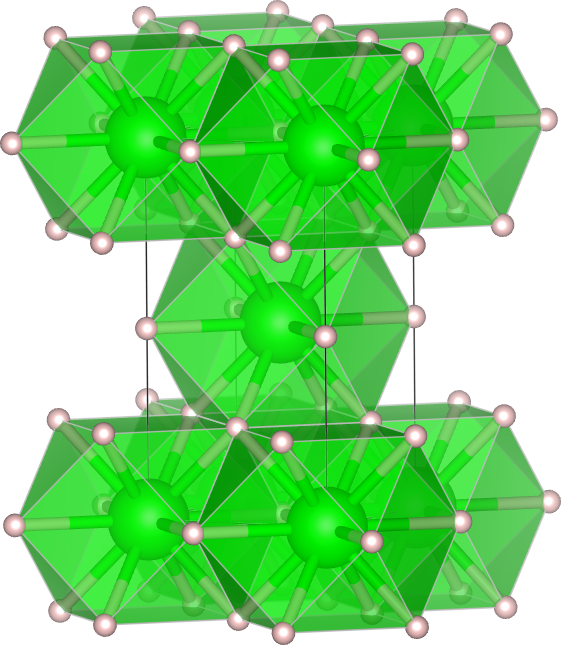} &
  \includegraphics[height=4cm]{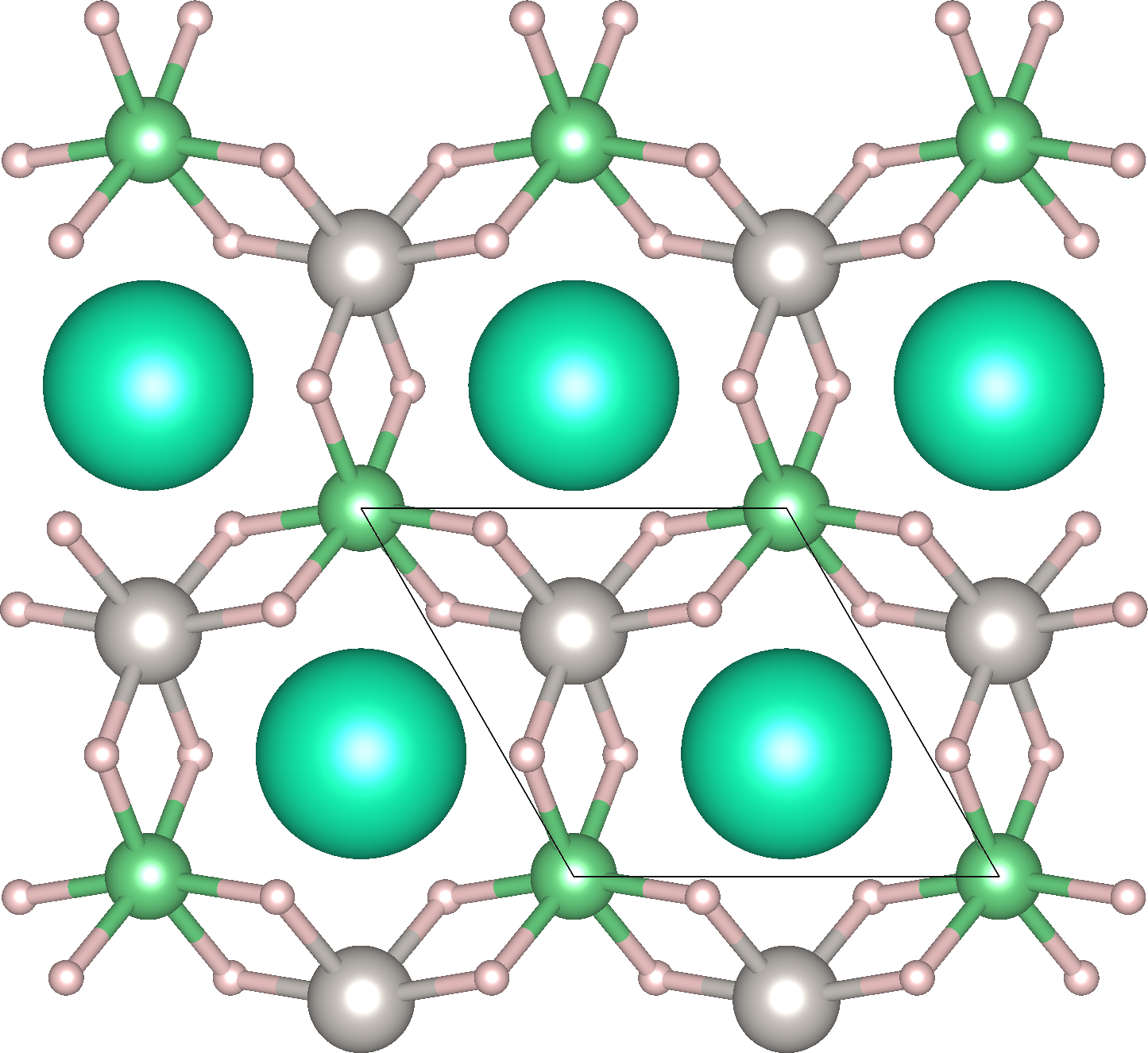}
\end{tabular}
\caption{Crystal structures of selected hydrides. The lines indicate the conventional cell. In the pictures 
H atoms are white, Li and Be green, Na yellow, Mg orange, K violet,  Cu dark blue, Zr green,  In pink, Cs cyan Pd and Pt grey, and Au orange.}
\label{fig:structures}
\end{figure*}

\begin{table}[htb]
    \caption{Superconducting transition temperatures calculated with the Allen-Dynes formula (\Tcad\ in K), logarithmic average of the phonon frequencies (\olog\ in K), average of the square of the phonon frequencies ($\omega^2$ in K), electron-phonon coupling constant ($\lambda$) and distance to the convex hull of stability ($E_\text{hull}$ in meV/atom).}
    \label{tab:propa}
\begin{tabular}{rrrrrrl}
  & \Tcad & \olog & $\omega^2$ & $\lambda$ & $E_\text{hull}$ & comment \\
  \hline \\[-2mm]
  \multicolumn{6}{c}{\bf SM$_2$-TM-H$_6$~\cite{10.48550/arxiv.2401.17024}} \\
  \ce{Li2CuH6} & 86.0  & 491 & 890 & 1.96 &  171 & \\ % H6Li2Cu_agm006188328 This one is not part of the public Alexandria; 
  % it was reported in https://arxiv.org/pdf/2401.17024.pdf and we had it calculated (as we have others not included in this table)
  % it also has a small instability in the interpolation around \Gamma
  \ce{Mg2PtH6} & 78.3  & 770  &  922 & 1.24 & 140 & \cite{Sanna_2024} \\ % H6Mg2Pt_agm002154264 
  \ce{Mg2PdH6} & 63.8  & 760  &  911 & 1.08 & 120 & \cite{Sanna_2024} \\ % H6Mg2Pd_agm002153626
  \ce{Mg2IrH6} & 59.4  & 634  & 1052 & 1.16 &  61 & \cite{Sanna_2024,10.48550/arxiv.2310.07562} \\% H6Mg2Ir_agm002153623
  \ce{Mg2RhH6} & 53.8  & 771  & 1132 & 0.96 &  30 & \cite{Sanna_2024} \\ % H6Mg2Rh_agm002153628
  \ce{Na2AgH6} & 46.1  & 511  & 1000 & 1.12 & 204 &  \\              % H6Na2Ag_agm006188406
  \ce{Al2MnH6} & 43.9  & 587  & 1014 & 1.00 &  68 &  \\              % H6Al2Mn_agm002374192
  \ce{Al2TcH6} & 43.1  & 484  &  964 & 1.11 &  53 &  \\              % H6Al2Tc_agm005682975 SUS
  \ce{Al2ReH6} & 36.4  & 505  &  994 & 0.98 &  55 & \\              % H6Al2Re_agm002374193
  \ce{Ca2AgH6} & 35.9  & 418  &  740 & 1.09 & 166 & \\              % H6Ca2Ag_agm005655392
  \ce{Ca2LiH6} & 33.0  & 433  &  706 & 1.01 & 226 & \\              % H6LiCa2_agm005655157
  \ce{Na2AuH6} & 31.6  & 632  & 1196 & 0.80 &  73 & \\              % H6Na2Au_agm002153637
  \ce{Na2CuH6} & 29.4  & 580  & 1122 & 0.81 & 135 & \\              % H6Na2Cu_agm002153639
  %\ce{Ga2ReH6} & 27.6  & 175  &  634 & 1.78 & 118 & \\              % H6Ga2Re_agm005683650 THIS HAS IMAGINARY FREQUENCIES
  \ce{Ga2RuH6} & 26.9  & 292  &  822 & 1.14 & 230 & \\              % H6Ga2Ru_agm005683457 SUS
  \ce{Mg2NiH6} & 22.7  & 921  & 1120 & 0.61 &  63 & \\              % H6Mg2Ni_agm002153625
  \ce{Ga2OsH6} & 20.9  & 262  &  841 & 1.04 & 179 & \\              % H6Ga2Os_agm005683659
%  \ce{Ga2MnH6} & 19.1  & 206  &  774 & 1.15 & 155 & \\              % H6MnGa2_agm005683343

  \\[-2mm]
  \multicolumn{6}{c}{\bf Perovskites} \\
  \ce{KInH3}  & 72.9  & 487  &  698 & 1.70 &  77 & \\              %  H3KIn_agm002490653
  \ce{Al4H}   & 30.4  & 430  &  546 & 0.97 & 136 & ~\cite{He2023a} \\ %   HAl4_agm006176992
  \ce{AlHgH3} & 28.4  & 403  &  854 & 0.96 & 278 & \\              % H3AlHg_agm002493244
  \ce{PbHgH3} & 25.8  & 462  &  808 & 0.85 & 420 & \\              % H3HgPb_agm002496201
  \ce{PbOsH3} & 23.2  & 292  &  722 & 1.03 & 369 & \\              % H3OsPb_agm002493550
  \ce{TiHMg3} & 21.8  & 196  &  362 & 1.31 & 370 & inverted \\     % HMg3Ti_agm002495459  

  \\[-2mm]
  \multicolumn{6}{c}{\bf Double Perovskites} \\
  \ce{K2InCuH6}  & 53.0  & 510  &  716 & 1.25 &  47 & \\ %  H6K2CuIn_agm004929677 ANTONIO
  \ce{K2LiCuH6}  & 47.1  & 470  &  719 & 1.21 & 116 & \\ %  H6LiK2Cu_agm002340583 ANTONIO
  \ce{Na2LiCuH6} & 45.9  & 666  &  950 & 0.95 & 102 & \\ % H6LiNa2Cu_agm002280446
  \ce{Na2SiPdH6} & 39.8  & 875  & 1062 & 0.77 &  84 & \\ % H6Na2SiPd_agm004984390
  \ce{Na2LiZnH6} & 37.8  & 519  &  763 & 0.98 & 146 & \\ % H6LiNa2Zn_agm002279787
  \ce{Na2GaRuH6} & 35.2  & 535  &  864 & 0.93 & 162 & \\ % H6Na2GaRu_agm004927065
  \ce{Cs2NaSnH6} & 34.5  & 324  &  588 & 1.27 &  67 & \\ % H6NaSnCs2_agm002153654
  \ce{K2AlHgH6}  & 32.5  & 349  &  701 & 1.15 &  86 & \\ %  H6AlK2Hg_agm004984734
  \ce{K2HgAuH6}  & 32.5  & 396  &  913 & 1.06 &  49 & \\ %  H6K2AuHg_agm004930514
  \ce{K2InAgH6}  & 31.7  & 660  &  860 & 0.79 &  32 & \\ %  H6K2AgIn_agm004930190
  \ce{Rb2AlHgH6} & 31.4  & 140  &  463 & 2.53 &  83 & \\ % H6AlRb2Hg_agm004984647
  \ce{Na2CdCuH6} & 26.0  & 740  &  986 & 0.69 &  63 & \\ % H6Na2CuCd_agm004928403
  \ce{Na2MgCuH6} & 24.7  & 251  &  707 & 1.20 &  92 & \\ % H6Na2MgCu_agm004929626
  \ce{Pb2CuRuH6} & 23.7  & 243  &  712 & 1.19 & 149 & \\ % H6CuRuPb2_agm004927040
  \ce{K2AlCdH6}  & 21.0  & 507  &  834 & 0.74 &  74 & \\ %  H6AlK2Cd_agm004984504

\end{tabular}
\end{table}

% SM2-TM-H6
\subsection{SM$_2$-TM-H$_6$}

Four elements of the SM$_2$-TM-H$_6$ family have already been discussed in Ref.~\cite{Sanna_2024}, namely \ce{Mg2RhH6}, \ce{Mg2IrH6}~\cite{10.48550/arxiv.2310.07562}, \ce{Mg2PdH6}, \ce{Mg2PtH6}, and a few more were proposed in Ref.~\cite{10.48550/arxiv.2401.17024}. This structure, depicted in \cref{fig:structures}a, is described by a face-centered cubic lattice with space group $Fm\Bar{3}m$. Here, the SM atoms are in the 8c, the TM atoms in the 4a, and the H atoms in the 24e Wyckoff positions. As we can see from~\cref{fig:structures}a, the H atoms form isolated octahedra around the TM atoms. All high-temperature superconductors we found within this family were perfectly symmetric and did not exhibit any deformation of the octahedra.

From the chemical point of view, in all our systems SM is a simple metal from groups IA, IIA, or IIIA while TM is a transition or a noble metal (see \cref{tab:propa}). It is simple to understand the chemical compositions in \cref{tab:propa} by a simple electron counting argument. The parent compound of this family, \ce{Mg2RuH6}~\cite{Kritikos1991,Huang1991}, is a stable semiconductor. Considering the standard oxidation states of +2 for Mg and -1 for H in hydrides, this leads to +2 for Ru, a standard oxidation state found in this metal (although less common than the +3 or +4 states). In total, this compound has 18 valence electrons that fill completely the electronic valence states. A superconducting state emerges when this compound is doped with electrons, either one per formula unit (e.g. \ce{Mg2IrH6}, \ce{Al2TcH6}, or \ce{Na2AuH6}) or two per formula unit (e.g. \ce{Mg2PtH6} or \ce{Ga2RuH6}).

From the electronic point of view, all compounds with SM being an alkali or an alkali earth are quite similar. The SM atom is fully ionized, so the top of the valence bands and the bottom of the conduction bands~\footnote{This nomenclature is relative to the band structure of the parent, semiconducting, compound.} are dominated by H and TM states. There are, of course, some changes to the dispersion of the bands that are system specific, leading to different density of states at the Fermi level, even for compounds that are isoelectronic. The situation is slightly more complicated when the SM is a group IIIA metal, as its states are not entirely ionized. This leads to an entirely different dispersion of the bands and therefore a different Fermi surface. 
In spite of this, the superconducting properties are in line with the ones of the IA and IIA-based compounds.

Concerning the phonon dispersions, the low energy modes are usually dominated by the TM atom, as expected by their larger mass, followed by a manifold composed mainly by the SM atom. However, in some cases where SM is a group IIIA metal we see a strong mixture of SM and TM states in the low energy range (e.g., in \ce{Al2TcH6} or in \ce{Al2ReH6}). The higher energy states, that contribute mostly to $\lambda$ are dominated by the vibrations of the hydrogen octahedra. The maximum phonon frequencies present in the spectrum can go from around 145~meV to almost 245~meV, depending on the chemical composition.

These compounds exhibit a wide range of values for \olog\ with the lowest being 292~K (\ce{Ga2RuH6}) to more than 770~K (\ce{Mg2PtH6} and \ce{Mg2RhH6}). On the other hand, the values of the electron-phonon coupling constant are typically around 1--1.2. Exceptions to this are \ce{Ga2RuH6}, that has a very soft mode at the $X$-point (that is also responsible for the abnormally low value of \olog~for this compound), and \ce{Mg2NiH6} that has the low value $\lambda=0.61$ (partially compensated by the high value of $\olog=921$~K). The combination of high \olog\ and relatively high $\lambda$ allows for dynamically stable crystals with the very high values of \Tc\ present in \cref{tab:propa}.

% Perovskites
\subsection{Perovskites}

Concerning the hydride perovskites, we find both normal (see \cref{fig:structures}b) and inverted systems. There are a series of hydride perovskites that have been synthesized experimentally~\cite{Ikeda2008}. Many of these are charge compensated, often combining an alkali and an alkali earth in positions 1a and 1b, such as \ce{LiBeH3}~\cite{Overhauser1987}, \ce{BaLiH3}~\cite{Messer1964} or \ce{KMgH3}~\cite{Maeland1993}. Not surprisingly, such systems are semiconducting. There are however, a few non-charge compensated systems that have been discovered, such as \ce{AlSrH3}~\cite{Kamegawa2014} (obtained through a high-pressure synthesis procedure) or \ce{SrPdH_{2.7}}~\cite{Bronger1994}.  There are also a few inverted (anti-)perovskites reported in the experimental literature, such as \ce{ChHM3} where M = Li, Na and Ch is a chalcogen~\cite{Gao2021} or \ce{InH_{0.8}Pd3}~\cite{Kohlmann2010}.

The band structure of \ce{KInH3} has a characteristic very dispersive, parabolic band centered at $R$ that crosses the Fermi level. 
All other hydride perovskites have very diverse electronic band structures and Fermi surfaces. We note due to the high Fermi velocity of the bands, the density of electronic states at the Fermi level is not extremely high in most cases. The low-energy phonons of the normal hydride perovskites are as expected dominated by the heavier atoms, but have a sizeable contribution to $\lambda$. In many of these systems, H modes are highly dispersive, and can be found around 30--180~meV. The inverted perovskite has a very different phonon disperion: In \ce{TiHMg3} the H-modes are lower in energy and are more dispersive, almost touching the low-energy Ti--Mg modes, and contribute significantly to the binding of the Cooper pairs.

Values of \olog\ for the normal perovskites are not particularly large, and are found in the range 40--60~meV. This is however compensated by rather large values of $\lambda$ that can reach 1.70 for \ce{KInH3}. In the inverted perovskite the situation is similar, but with even lower values of \olog. Finally, we would like to mention \ce{Al4H}, a material proposed in Ref.~\cite{He2023a} where Al occupies both the 1b and 3d Wyckoff positions of the cubic perovskite structure.

% Double perovskites
\subsection{Double perovskites}
In view of the number of normal perovskites we found, it is not surprising that several double perovskite systems (see \cref{fig:structures}c) also appeared in our survey. From \cref{tab:propa} we see that the 8c position (the counterpart of the 1a position of the ternary perovskites) is mostly occupied by an alkali (with the exception of \ce{Pb2CuRuH6}), while the 4a and 4b positions include two cations, often combining a SM with a TM. Of course, the quaternary system allows for a much larger number of chemical compositions than the ternary one.

To our knowledge there are no experimentally synthesized hydride double perovskites. However, these systems have been proposed theoretically as high-temperature ferromagnetic semiconductors~\cite{Jia2023} and the effect of strain on the hydrogen storage characteristics in \ce{K2NaAlH6} has been studied in Ref.~\onlinecite{Baaddi2023}.

What is perhaps surprising is the diversity of electronic band structures and of Fermi surfaces for the double perovskites of \cref{tab:propa}. In all cases with an alkali in 8c position, these atoms are fully ionized and do not contribute to the density of states at the Fermi level. The Fermi level, instead, contains states of H and of the cations in positions 4a and 4b. The chemical diversity of these ions, with some having $d$ states and others not, leads to different band structures close to the Fermi level. From this we can conclude that the detailed shape of the Fermi surface or the Fermi velocity are not determinant factors for the superconducting properties of double perovskite hydrides, but only the contribution of H to the density of states at the Fermi level~\cite{belli2021strong}.
The same holds true for the phonon band structure. Although the cations in the 4a and 4b positions are normally responsible for the acoustic and low-lying optical modes, the higher lying optical modes show a variety of dispersions and frequencies. The highest optical mode is usually in the range of 135--175~K.

We can see from \cref{tab:propa} that values of $\lambda$ are between 0.7--1.25, while \olog~is between 250 and 875~K. These compounds show extremely well that the highest \Tc\ is obtained through the synergy between $\lambda$ and \olog, and not by extreme values of either quantity. We emphasize that in none of the normal, inverted, or double perovskites we encounter any rotation or tilting of the octahedra that are so common in oxide perovskites.

\begin{table}[htb]
    \caption{Continuation of \cref{tab:propa}.}
    \label{tab:propb}
\begin{tabular}{rrrrrrl}
  & \Tcad & \olog & $\omega^2$ & $\lambda$ & $E_\text{hull}$ & comment \\
  \hline \\[-2mm]
  \ce{Be8H}     & 65.9  & 588  &  718 & 1.34 & 202 & \cite{He2023b} \\ %      HBe8_agm006190471 
  \\[-2mm]
  \ce{Mg2SrPtH8}& 49.4  & 568 &  925 & 1.10 & 117 &              \\ % H8Mg2SrPt_agm002280639
  \ce{Mg2SrPdH8}& 49.1  & 652  &  925 & 1.00 &  93 &              \\ % H8Mg2SrPd_agm002280638 ANTONIO
  \ce{Mg2BaPdH8}& 41.6  & 646 &  939 & 0.92 &  63 &              \\ % H8Mg2PdBa_agm002280631
  \ce{Mg2BaPtH8}& 38.4  & 590  &  961 & 0.92 & 102 &              \\ % H8Mg2BaPt_agm002280563
  \\[-2mm]
  \ce{Na2AuH4}  & 43.8  & 394  &  598 & 1.31 &  35 &              \\ %   H4Na2Au_agm002153449 ANTONIO
  \ce{LiPdH2}   & 38.6  & 532  &  849 & 0.98 & 108 & \cite{newAdvMat} \\ %    H2LiPd_agm001945535 ANTONIO % WE ALREADY DISCUSS THIS ONE IN THE DB PAPER
  \ce{NaPdH3}   & 29.3  & 215  &  722 & 1.56 & 193 & \\              %    H3NaPd_agm003068400
  \ce{Na3Pd2H9} & 25.3  & 516  & 1031 & 0.80 &  86 & \\              %  H9Na3Pd2_agm005611460
  \ce{Li2AuH2}  & 20.8  & 368  &  762 & 0.86 &  59 & \\              %   H2Li2Au_agm002153052
  \\[-2mm]
  \ce{Li4BeH5}  & 37.2  & 465  &  716 & 1.04 & 154 & \\              %   H5Li4Be_agm003659872 ANTONIO
  \\[-2mm]
  \ce{PdH}      & 34.9  & 343  &  429 & 1.24 &   8 & \cite{Stritzker1972,Errea2013} \\ %       HPd_agm003157455
  \ce{InH}      & 31.2  & 312  &  724 & 1.20 & 223 & \\              %       HIn_agm002078382
  \ce{ZrH3}     & 25.8  & 363  &  748 & 0.97 & 105 & \cite{PRM2023-ZrH3-isotope} \\ %      H3Zr_agm001282320
  \ce{InH3}     & 20.6  & 218 &  700 & 1.17 & 286 &              \\ %      H3In_agm002153249
  \ce{Zr2HfH8}  & 25.1  & 266 &  663 & 1.16 &  94 &              \\ %   H8Zr2Hf_agm005945812 beautiful hexagons
  \\[-2mm]
  \ce{KBePtH6}  & 39.9  & 371 &  764 & 1.27 & 138 &              \\ %   H6BeKPt_agm005096310
  \ce{RbBePtH6} & 37.9  & 379 &  767 & 1.21 & 135 &              \\ %  H6BeRbPt_agm005096259
  \ce{CsBePtH6} & 30.9  & 370 &  787 & 1.07 & 130 &              \\ %  H6BeCsPt_agm005096309
%  \ce{CsBePdH6} & 33.8  & 158 &  554 & 2.43 & 104 &              \\ %  H6BePdCs_agm005098789 Very likely imaginary
  \\[-2mm]
  \ce{Li2CuGaH6}& 37.9  & 527 &  891 & 0.97 & 133 &              \\ % H6Li2CuGa_agm004959278 not a double perovskite, but with octahedra
  \ce{K2AgCdH6} & 27.5  & 422 &  775 & 0.92 &  75 &              \\ %  H6K2AgCd_agm004928433
  \ce{Na2PdIrH6}& 23.5  & 456 &  736 & 0.82 &  93 &              \\ % H6Na2PdIr_agm004930563 Na2IrH6 layers intercalated with Pd
\end{tabular}
\end{table}

% Others
\subsection{Other compounds}

From the compounds in \cref{tab:propb}, the one with highest predicted \Tc\ is \ce{Be8H}. This material, proposed as a high-temperature superconductor in Ref.~\onlinecite{He2023b} is composed by double triangular layers of Be intercalated with triangular layers of H. It has a complicated band structure, with different bands crossing the Fermi energy. Interestingly, the large value of $\lambda=1.34$ is mainly due to the lowest-lying optical modes of Be, although with a non-negligible contribution of the three high-frequency modes composed mostly of H vibrations.

We then encounter four closely related systems, specifically \ce{Mg2\{Sr,Ba\}\{Pt, Pd\}H8} with predicted transition temperatures of around 49~K (for the Sr compounds) and 40~K for the Ba compounds. These crystallize on a trigonal phase with alternating layers of \ce{Mg2\{Pt,Pd\}H6} and \ce{\{Sr, Ba\}H2}. We find two bands crossing the Fermi surface, with one of the crossings (at the H point) exactly at the Fermi energy. Due to the difference in atomic masses, the phonon frequencies are well separated, with the acoustic and lowest-lying optical branches with \ce{\{Sr,Ba\}} and \ce{\{Pt, Pd\}} character, followed by modes with mostly Mg character. Finally, the H vibrations form two separated manifolds at higher energy spanning frequencies from 40 to 140~meV. The electron-phonon coupling constants for these materials are around $\lambda=1$, but \olog\ has large values of around 600--650~K.

With transition temperatures between 20 and 44~K we find a series of systems with similar chemistry, specifically with compositions involving one light alkali metal (Li or Na) and one noble metal (Au or Pd, see examples in \cref{fig:structures}d--f). The one with the highest transition temperature is \ce{Na2AuH4}, a tetragonal compound with alternating layers of Na and \ce{AuH4}. In the latter compound, the H atoms form the vertices of isolated squares aligned with the direction of the layer, with the Au atoms at the center of the squares. There are two bands crossing the Fermi surface, with a Dirac band crossing at the Fermi energy at a point between X and $\Sigma$. The relatively high \Tc\ is mostly due to the electrons with the lowest-lying H-modes at around 40~meV.

The next compound, \ce{LiPdH2} has the delafossite structure with \ce{PdH2} layers intercalated with Li (see \cref{fig:structures}d). This material, with a predicted \Tc\ of almost 40~K has already been discussed in Ref.~\onlinecite{newAdvMat}. Next come two Na-Pd hydrides, with very different stoichiometries, structures, and superconducting properties, but with similar \Tc. The first, \ce{NaPdH3}, is cubic, with sharing \ce{PdH6} octahedra tilted with respect to each other, with all H atoms shared between two octahedra. The second, \ce{Na3Pd2H9} has aligned \ce{PdH6} octahedra that form chains through the sharing of two H per octahedron (see \cref{fig:structures}e). These two compounds exhibit very different band structures: the first has many bands crossing the Fermi surface, leading to a complicated Fermi surface; the second has two bands at the Fermi level, with several crossings at, or near, the Fermi energy. Both compounds have phonons arranged in three well separated manifolds with comparable energies, with the acoustic and low-lying optical branches with Pd--Na character, and the two higher lying manifolds due to H-vibrations of the octahedra.  Finally, \ce{Li2AuH2} has a very interesting structure, with \ce{AuH2} linear units alternating with layers of Li. The values of $\lambda=0.86$ and $\olog=368$~K are relatively modest for the hydrides in this list, leading to a \Tc\ around 20~K.

The structure of \ce{Li4BeH5} is very interesting, with the Li atoms forming a snub-square lattice~\cite{Wang2023} when viewed from the $c$-axis. There are two electronic bands crossing the Fermi surface, one with high Fermi velocity while the other forms a pocket around the $M$-point. The Fermi level lies on a steep shoulder of the density-of-states, suggesting the \Tc\ could be increased by electron doping. The phonon branches contributing the most to $\lambda$ are the lower-lying modes, where one sees a strong mixture of Li, Be, and H modes.

Next in \cref{tab:propb} come a series of simple transition metal hydrides with \Tc\ in the range of 20--35~K. \ce{PdH} is a well-known superconductor~\cite{Stritzker1972} where anharmonic effects have a huge effect in decreasing the transition temperature~\cite{Errea2013}. On the other hand, \ce{ZrH3} (see \cref{fig:structures}g) has been recently reported with a $\Tc=11.6$~K~\cite{PRM2023-ZrH3-isotope}. We also find two indium hydrides: The rather thermodynamically unstable \ce{InH} has very simple hexagonal structure with two atoms in the unit cell. There are two very dispersive bands crossing the Fermi surface, leading to a relatively low density of states at the Fermi surface. Indium and hydrogen phonons are well separated, with the three acoustic (hydrogen) and the first second optical (indium) branches contributing most to $\lambda$. Finally \ce{Zr2HfH8} has a curious crystal structure that has the form of a Triakis triangular tiling, with \{Zr,Hf\} in the hexagonal positions, when viewed from the $a$-axis. Of course, due to similarity between Zr and Hf we can expect a certain amount of alloying in the cation positions. The Fermi surface is quite complicated, with many bands crossing the Fermi surface, and the Fermi level lies on steep shoulder of the density of states. Most of the value of $\lambda=1.16$ comes from the lowest manifold of phonon bands composed exclusively of Zr and Hf vibrations and from the lowest hydrogen vibrations, while the higher optical hydrogen modes couple very weakly to the electrons.

The three compounds \ce{\{K, Rb, Cs\}BePtH8} have a hexagonal structure, with \ce{BePtH8} layers intercalated by alkali layers (see \cref{fig:structures}h). The band structure close to the Fermi level is essentially determined by the \ce{BePtH8} layers. These are characteristic of an indirect band gap semiconductor with heavy holes and light electrons. The extra electron donated by the alkali atom half-fills the conduction band leading to a metallic ground-state. The strong electron-phonon coupling constants $\lambda=1.07$--1.27 are mainly due to vibrations of the heavy atoms and the lowest lying hybridized Be--H modes. We also see that \olog\ is basically independent of the alkali, while $\lambda$ decreases going from K to Cs, leading to a decrease of \Tc.

Finally, we find three systems, \ce{Li2CuGaH6}, \ce{K2AgCdH6} and \ce{Na2PdIrH6} with similar chemical compositions, but different crystal structures and electronic and phononic band structures. In spite of this, they all have similar values of $\lambda$ and \olog, leading to \Tc\ between 23 and 37~K.

\subsection{Statistical analysis}\label{sec:statistics}

\begin{figure}
\includegraphics[width=8cm]{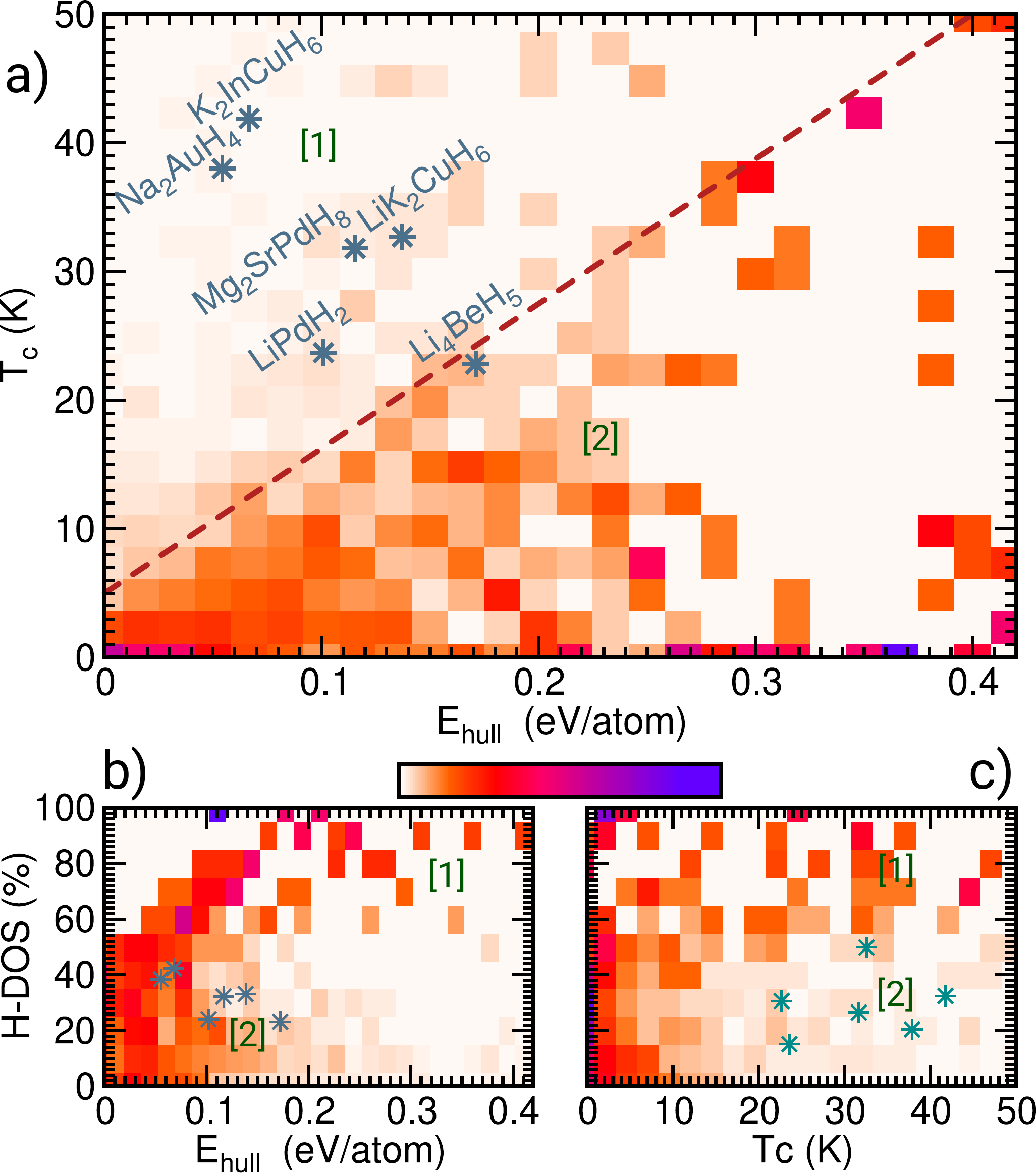}
\caption{Analysis of superconducting hydrides:. a) Histogram of \Tc\ as a function of the energy from the hull. The dashed line is a guide to the eye marking a threshold in critical temperature separating high from low probability regions. The histogram is normalized column-wise and a color-bar (white to blue) indicates the normalized histogram height. b) Relation between energy from the hull and the hydrogen contribution to the density of states at the Fermi level. c) Relation between \Tc\ and the hydrogen contribution to the density of states at the Fermi level. The histograms in b) and c) are normalized row-wise. Stars show some of the materials discussed in \cref{sec:selectedHydrides}.}
\label{fig:Tc_energy_and_Hprojections}
\end{figure}

These examples show that there is not a unique or simple recipe linked to strong superconductivity in hydrides at ambient pressures. However, we can see some patterns from a statistical analysis of our data-set. In \cref{fig:Tc_energy_and_Hprojections}a we analyze the correlation between the thermodynamic stability and \Tc. This figure is obtained by grouping the materials according to their energy from the hull and, for each group, computing the histogram of \Tc\ values (i.e. the number of materials in the figure is normalized column-wise). The figure shows that, while there is not a direct proportionality between \Tc\ and thermodynamic stability, the chance of finding high-\Tc\ outliers increases significantly with the distance from the hull.  We identify two regions (separated by a dashed line): a high-\Tc\ zone where materials are rare (``[1]'' in the figure) and a low-\Tc~zone with high density of materials (``[2]'' in figure). Ideally we would be interested in high-\Tc\ stable materials, but this set appears to be rather empty. Therefore the sparse region of high-stability and high-\Tc\ is the most promising search domain. 

A second aspect that emerges from our analysis is the complex relation between hydrogen and superconductivity. As shown in \cref{fig:Tc_energy_and_Hprojections}c there is a correlation between high hydrogen content at the Fermi level (as measured by computing the H fraction of the density of states) and \Tc. High hydrogen content is linked (as it should be~\cite{Ashcroft2004,belli2021strong}) to higher values of \Tc\ (region ``[1]'' in \cref{fig:Tc_energy_and_Hprojections}c.  Unfortunately, the high hydrogen content also correlates with thermodynamic instability. As shown in \cref{fig:Tc_energy_and_Hprojections}b, by increasing the H content the center of mass of the material distribution shifts towards higher energy (region ``[1]'' of the plot).  Note that the data is normalized row-wise. From this figure it is clear that the region of interest is in materials for which the Fermi density of states is dominated by the host chemical composition and the H contribution is less than 50\% of the total. This region is indicated by the label ``[2]'' in \cref{fig:Tc_energy_and_Hprojections}b,c.

\subsection{Analysis of selected hydrides}
\label{sec:selectedHydrides}

\begin{table}
\caption{Electron-phonon coupling constant $\lambda$, logarithmic average of the phonon frequencies \olog\ (in K), density of states at the Fermi level \dosef\ (in (eV cell)$^{-1}$), the H content at the Fermi level (\%), the Coulomb potential $\mu$, transition temperatures \Tc\ (in K) and superconducting gaps $\Delta$ (in meV) for the twelve compounds reported here. Calculations were performed with the isotropic Eliashberg equations including full ab initio Coulomb interactions.}
\label{tab:selected}
\begin{center}
\begin{tabular}{ l r r r r r r r}
              & $\lambda$ & \olog & $N_F$  & $H$\%  & $\mu$  & \Tc   & $\Delta$ \\     \hline\\[-2mm]
\ce{Mg2RhH6}  & 1.3       & 766    & 2.32   & 18.4   & 0.45   & 48.5  & 8.4      \\
\ce{Mg2PdH6}  & 1.1       & 754    & 0.80   & 28.3   & 0.17   & 66.5  & 11.2     \\
\ce{Mg2IrH6}  & 2.13      & 581  & 2.90   & 15.2   & 0.58   & 77.0  & 16.3     \\
\ce{Mg2PtH6}  & 1.4       & 696   & 1.03   & 22.6   & 0.20   & 80.4  & 15.0     \\
\ce{K2InCuH6} & 1.14      & 448  & 1.38   & 32.2   & 0.26   & 41.9  &  7.0     \\
\ce{K2LiCuH6} & 1.28      & 354  & 2.34   & 49.6   & 0.35   & 32.7  &  5.6     \\
\ce{Mg2SrPdH8}& 0.84      & 608  & 1.10   & 26.3   & 0.19   & 31.8  &  5.1     \\
\ce{Na2AuH4}  & 1.28      & 391  & 1.09   & 20.2   & 0.27   & 38.0  &  6.6     \\
\ce{LiPdH2}   & 0.80      & 454  & 0.53   & 14.9   & 0.16   & 23.7  &  3.8     \\
\ce{Li4BeH5}  & 0.95      & 460  & 1.05   & 30.4   & 0.31   & 22.8  &  3.7     \\
\end{tabular}
\end{center}
\end{table}

Among the superconducting hydrides predicted in our high throughput search we have selected a small set of compounds and investigated these with higher accuracy and in further detail. There are two main problems we want to address here: (i)~The high density of states at the Fermi level of these compounds, and the fact that the Fermi level is often at a steep shoulder or a peak, which makes the convergence with respect to smearing and $k$-points very complicated. (ii)~The replacement of the Coulomb interaction by the number $\mu^*$ that is given a rather arbitrary value (usually around 0.10) is hardly justified for many of the high-\Tc\ hydrides~\cite{FloresLivas_SH3_EPJB2016,Akashi_SH3_PRB2015,FloresLivasReveiwHydrides2020}.
Instead, the full energy dependence of the Coulomb term should be maintained in the calculation of the superconducting properties.

The list of compounds that we selected, together with some of their properties are collected in \cref{tab:selected}, which includes four materials \ce{Mg2XH6} with X=\ce{Rh},\ce{Pd},\ce{Ir},\ce{Pt} already discussed in Ref.~\cite{Sanna_2024}. These are outlier materials with respect to the general trends discussed in the previous section, close to the hull of stability and with high \Tc. %Their position in the plots of Fig.~\ref{fig:Tc_energy_and_Hprojections} is marked by a blue star.
First we observe that the calculated \Tc s reported in \cref{tab:selected} are in most cases lower than that listed in \cref{tab:propa} and \cref{tab:propb}. In some cases, such as \ce{LiPdH2} and \ce{Mg2SrPdH8}, the difference is related to changes in the estimated e-ph coupling parameters (owing to the increased numerical accuracy), however in most of the cases the e-ph coupling is confirmed and the lower \Tc\ stems from a strong Coulomb repulsion, which here we compute from first principles.

\begin{figure*}
%\begin{flushleft}
%a)\vspace{-0.7cm}
%\end{flushleft}
%\includegraphics[width=\textwidth]{img/phbands.pdf}
%\begin{flushleft}
%b)\vspace{-0.5cm}
%\end{flushleft}
\includegraphics[width=\textwidth]{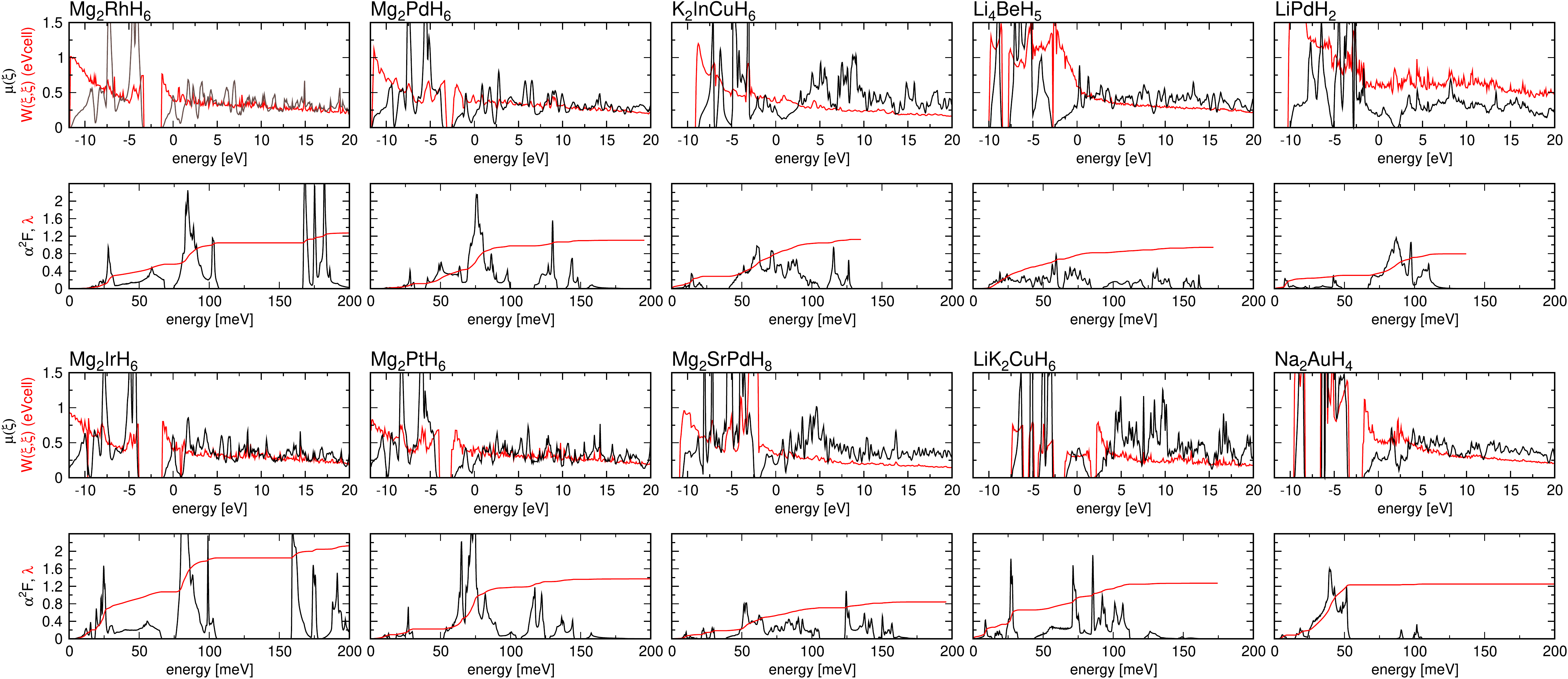}
\caption{
%Phonons and coupling parameters of selected hydrides. a) Phonon bands and density of states. Colors indicate the atomic projections of the phonon modes according to the dynamical matrix.  Hydrogen modes are shown in dark-red. b) 
Coupling parameters entering the isotropic Eliashberg equations~\cite{Pellegrini_SimpEliashberg_JOPM2022,Adavydov_PRB2020}. The quantity $W(\xi,\xi')$ is the Coulomb interaction computed in the random phase approximation,  $\mu(\xi):=W(\xi,\xi)N(\xi)$ is the product of $W$ and the density of electronic states. Finally, $\alpha^2F(\omega)$ is the Eliashberg electron-phonon function, $\lambda(\omega):=2\int_0^{\omega}\frac{\alpha^2F(w)}{w}dw$ is the coupling integration curve~\cite{FloresLivasReveiwHydrides2020}. The colors of the plotted lines in each panel correspond to the labels on the vertical axes.}
\label{fig:phbands}
\end{figure*}

The hydrogen contribution to the pairing is evident from the phonon band structure and electron phonon coupling functions reported in Fig.~\ref{fig:phbands}. In fact, more than 50\% of the integrated coupling constant $\lambda$ originates from pure-hydrogen phonon modes, which as we discussed above, tend to be well separated from the vibrational modes of the host elements. The only exception to this is \ce{Li4BeH5}, where we see a low energy region of mix-modes. In stable materials at low pressure hydrogen bonds are usually completely saturated, therefore hydrogen modes have no coupling with states at the Fermi level. What makes these hydrides quite special is that hydrogen states contribute significantly to the Fermi density of states (of the order of 25\% according to atomic projections), furthermore they are not localized but strongly overlapping and hybridizing with the states of the host similarly to what occurs in high pressure superconducting hydrides~\cite{FloresLivasReveiwHydrides2020}. 

The screened coulomb interaction $W$ (red curves in the panels in the first and third rows of Fig.~\ref{fig:phbands}) is computed in the random phase approximation which is expected to be quite accurate for this class of materials~\cite{Pellegrini_KO_PRB2023}. The diagonal part of the $W$-function is relatively featureless, showing a nearly monotonous decrease with the band energy. This behaviour, reminiscent of free electron systems, indicates that electronic states are well screened~\cite{Massidda_Sanna_SUST,Pellegrini_KO_PRB2023}. However the strength and effectiveness of the Coulomb repulsion is strongly affected by the density of states which shows a complex material dependence. The most peculiar cases are the \ce{Mg2IrH6} and \ce{Mg2RhH6} systems where the Fermi level is pinned to a peak in the density of states, leading to a complicated (and quite poor) Coulomb renormalization mechanism~\cite{ScalSchrWilk_strongCouplingI_PRB1966}. To a lesser extent a peak in the density of states also causes \ce{LiK2CuH6} to be characterized by a strong Coulomb repulsion. The other materials in the set are less problematic. 

\section{Conclusion}

We have presented a machine-learning assisted approach to search for superconducting hydrides under ambient pressure within an extensive dataset comprising over 150\,000 compounds. Our investigation yielded around 50 systems with transition temperatures surpassing 20~K, and reaching above 70~K. These systems have a large variety of crystal structures, and geometrical characteristics. However, we find a consistent chemical composition in the majority of these systems, which combines alkali or alkali-earth elements with noble metals. 

On the down side we observe that most of the systems we find, while being dynamically stable, are not on the hull of thermodynamic stability (while being close to it). This points to the fact that the experimental synthesis of our predicted compounds may require conditions beyond ambient equilibrium and skilled material engineering strategies. 

A statistical analysis of our predictions, as well as in-depth theoretical characterization for a set of selected materials, provide a reasonably clear picture of the superconducting mechanism in hydrides at room pressure. The crucial aspect is that hydrogen states are present at the Fermi level, strongly hybdridized with those of the host. These are delocalized (non-molecular) metallic states that are well screened and lead to large values of the density of states. In this situation, both medium or high-energy hydrogen phonon modes and low energy host modes contribute to the coupling $\lambda$. The final value of \Tc\ is a complex interplay of large density of states, correct phonon energies, high hydrogen contribution to the Fermi density of states and low Coulomb repulsion. In a statistical sense, we observe that there is a direct correlation between low thermodynamic stability, high hydrogen content and high \Tc. This observation suggests a promising avenue for future experimental investigations into high-temperature superconductivity within hydrides at ambient pressure.

\section{Methods}

All density-functional calculations are performed using the versions 6.8 and 7.1 of \textsc{quantum espresso}~\cite{Giannozzi2009,Giannozzi2017} with the Perdew-Burke-Ernzerhof (PBE) for solids (PBEsol)~\cite{pbesol} generalized gradient approximation. 

Geometry optimizations are performed using uniform $\Gamma$-centered $k$-point grids with a density of 3000~$k$-points per reciprocal atom. If this results in an odd number of $k$-points in a given direction, the next even number is used instead. Convergence thresholds for energies, forces and stresses are set to $1\times10^{-8}$\,a.u., $1\times10^{-6}$\,a.u., and $5\times10^{-2}$~kbar, respectively. For the electron-phonon coupling we use a double-grid technique, with the $k$-grid used in the lattice optimization doubled in every direction as the coarse grid, and a $k$-grid quadrupled in each direction as the fine grid. For the $q$-sampling of the phonons we use half of the $k$-point grid described above. The double $\delta$-integration to obtain the Eliashberg function is performed with a Methfessel–Paxton smearing of 0.05~Ry.

Distances to the convex hull were computed using the PBEsol approximation~\cite{pbesol}. We considered all pertinent materials available in the Alexandria database~\cite{schmidtCrystalGraphAttention2021,10.1002/adma.202210788} and re-optimized them following Ref.~\onlinecite{dataset}.

For the machine learning part, we trained the \textsc{alignn}~\cite{Choudhary2021} model using as targets, simultaneously, $\lambda$, \olog\ and \Tc\, with the error for each property weighted equally, as these are the choices yielding the best results. We used the default hyperparameters. The final models can be downloaded from \url{https://github.com/hyllios/utils/tree/main/models}.

A few superconductivity simulations were performed with isotropic Eliashberg theory~\cite{Pellegrini_SimpEliashberg_JOPM2022,Adavydov_PRB2020} including full ab-initio Coulomb interactions. The Eliashberg $\alpha^2F$ function was computed using quantum espresso~\cite{Giannozzi2009,Giannozzi2017} and density functional perturbation theory~\cite{Baroni_DFPT_RMP2001,PhysRevLett.58.1861}. A converged momentum integration was performed by mapping the calculated electron phonon matrix elements to a random set of 60\,000 k-points accumulated on the Fermi surface~\cite{Sanna_NbSe2_npjQM2022,Marques_SCDFT_PRB2005}.  All calculations were done using PBEsol pseudo potentials from the strict set of PseudoDojo~\cite{vanSetten2018pseudodojo} and a 100~Ry energy cutoff for the planewave expansion. A $16^3$($8^3$) $k$-grid($q$-grid) was used for the zone sampling in calculating the dynamical matrices of \ce{Mg2XH6} and \ce{LiPdH2}, a $12^3$($6^3$) $k$-grid($q$-grid) was used for the other systems in ~\cref{tab:selected}. 
Screening was computed~\cite{ElkCode} in the random phase approximation which is expected to be sufficiently accurate for this class of materials~\cite{Pellegrini_KO_PRB2023}. The response function was calculated with a $G$-vector cutoff of 3~atomic units and an energy window of 2~Hartrees. The $k$-grids for momentum integration of the Coulomb interaction were converged to 10\% in the value of $\mu$, and ranged from $5^3$ in \ce{Mg2SrPdH8} to $12^3$ in \ce{LiPdH2}.

\section{Data availability statement}

All relevant data is present in the manuscript and in the Supplementary
Information.

\section{Acknowledgements}
T.F.T.C acknowledges financial support from FCT - Fundação para a Ciência e Tecnologia, I.P. through the projects UIDB/04564/2020, UIDP/04564/2020 and CEECINST/00152/2018/CP1570/CT0006, with DOI identifiers 10.54499/UIDB/04564/2020, 10.54499/UIDP/04564/2020, and 10.54499/CEECINST/00152/2018/CP1570/CT0006, respectively. Computational resources provided by UC-LCA, funded by FCT I.P. under the project Advanced Computing Project 2022.15822.CPCA. M.A.L.M. acknowledges partial funding from Horizon Europe MSCA Doctoral network grant n.101073486, EUSpecLab, funded by the European Union, and from the Keele Foundation through the SuperC collaboration. 
Y.-W.F. and I.E. received funding from the European Research Council (ERC) under the European Union’s Horizon 2020 research and innovation programme (Grant Agreement No. 802533) and acknowledge PRACE for awarding access to the EuroHPC supercomputer LUMI located in CSC's data center in Kajaani, Finland through EuroHPC Joint Undertaking (EHPC-REG-2022R03-090). I.E. also acknowledges funding from the Spanish Ministry of Science and Innovation (Grant No. PID2022-142861NA-I00) and the Department of Education, Universities and Research of the Basque Government and the University of the Basque Country (Grant No. IT1527-22). The authors acknowledge enlightening discussions with the partners of the SuperC collaboration.

\section{Author Contributions}

T.F.T.C. and M.A.L.M. performed the machine learning prediction and the preliminary analysis of the superconducting properties. A.S. performed the statistical analysis and the analysis of selected hydrides. All authors contributed to designing the research, interpreting the results and writing of the manuscript.

\section{Competing  Interests}

The authors declare that they have no competing interests.

\bibliography{main.bib}

\end{document}